\documentstyle[11pt]{article} 

\def \inbar{\vrule height1.5ex width.4pt depth0pt}
\def \xC{\relax\hbox{\kern.25em$\inbar\kern-.3em{\rm C}$}}
\def \xR{\relax{\rm I\kern-.18em R}}
\newcommand{\R}{\xR}

\newcommand{\xbe}{\begin{equation}}
\newcommand{\be}{\begin{equation}}
\newcommand{\xee}{\end{equation}}
\newcommand{\ee}{\end{equation}}
\newcommand{\xbea}{\begin{eqnarray}}
\newcommand{\bea}{\begin{eqnarray}}
\newcommand{\xeea}{\end{eqnarray}}
\newcommand{\eea}{\end{eqnarray}}

\newcommand{\nn}{\nonumber}

\newcommand{\kt}{\rangle}

\newcommand{\br}{\langle}

\newcommand{\cun}{\mbox{\footnotesize${\cal N}$}}

\newcommand{\ed}{\end{document}}

\begin{document}

\title{Geometric Phases, Symmetries of Dynamical Invariants, and Exact Solution of 
the Schr\"odinger Equation}
\author{Ali Mostafazadeh\thanks{E-mail address: 
amostafazadeh@ku.edu.tr}\\ \\
Department of Mathematics, Ko\c{c} University,\\
Rumelifeneri Yolu, 80910 Sariyer, Istanbul, TURKEY}
\date{ }
\maketitle

\begin{abstract}
We introduce the notion of the geometrically equivalent quantum systems (GEQS) as 
quantum systems that lead to the same geometric phases for a given complete set of 
initial state vectors. We give a characterization of the GEQS. These systems have a common
dynamical invariant, and their Hamiltonians and evolution operators are related by symmetry 
transformations of the invariant. If the invariant is $T$-periodic, the corresponding class of 
GEQS includes a system with a $T$-periodic Hamiltonian. We apply our general results to study the 
classes of GEQS that include a system with a cranked Hamiltonian 
$H(t)=e^{-iKt}H_0e^{iKt}$. We show that the cranking operator $K$ also belongs 
to this class. Hence, in spite of the fact that it is time-independent, it leads to nontrivial 
cyclic evolutions and geometric phases. Our analysis allows for an explicit construction of a 
complete set of nonstationary cyclic states of any time-independent simple harmonic oscillator. The period 
of these cyclic states is half the characteristic period of the oscillator.
\end{abstract}
\vspace{2mm}


\baselineskip=24pt

\section{Introduction}

History of modern physics is full of happy surprises. Among the latest of these
is the discovery of Berry's adiabatic geometric phase 
\cite{mead,berry-1984,review,bohm-qm,nova}. This discovery has led physicists to 
reconsider a number of fundamental as well as practical aspects of 
quantum mechanics. The importance of Berry's findings and their impact on 
various areas of physics have naturally resulted in the interest in the 
generalizations of geometric phases. One of the most significant contributions in 
this direction is the nonadiabatic generalization of Berry's phase due to
Aharonov and Anandan \cite{aa}. This generalization employs a geometric 
picture of quantum dynamics and shows that the nonadiabatic geometric phase 
can be defined for any closed curve in the space of (pure) quantum states.
Moreover, the geometric phase associated with the evolution of a pure state 
only depends on the path traced by the evolving state in the state space. In 
other words, different Hamiltonians leading to the same path define the same 
geometric phase.\footnote{This is usually demonstrated by showing that a shift of the 
Hamiltonian $H(t)$ by a multiple of the identity operator $I$, i.e., $H(t)\to H'(t)
=H(t)+f(t)I$, leaves the geometric phases invariant \cite{bohm-qm}.This shift is related
to a global phase transformation of the Hilbert space. It is in a sense a trivial kinematic 
symmetry transformation \cite{nova}.} The purpose of this article is to address the 
problem of the characterization of all the quantum systems (with a common Hilbert 
space) that lead to the same geometric phases for a complete set of initial state vectors.
We will term such systems `{\em geometrically equivalent},' and determine the Hamiltonian
$\tilde H(t)$ for a general quantum system that is geometrically equivalent to a given system 
with Hamiltonian $H(t)$. 

Our main results are:
	\begin{itemize}
	\item[1.] This problem is directly linked with the symmetries of 
	dynamical invariants associated with the Hamiltonian $H(t)$;
	\item[2.] The evolution operator $\tilde U(t)$ for $\tilde H(t)$
	can be obtained from the evolution operator $U(t)$ for $H(t)$. 
	\item[3.] Each class of geometrically equivalent quantum systems
	includes a system whose dynamical and geometric phases cancel
	each other, i.e., their total phase is unity.\footnote{Here we mean the
	phases associated with the eigenvectors of the common dynamical
	invariant.} In particular, if the corresponding dynamical invariant 
	is $T$-periodic, the class includes a system with a $T$-periodic 
	Hamiltonian $H_*(t)$. Moreover, the evolution operator $U_*(t)$
	corresponding to $H_*(t)$ satisfies $U_*(T)=1$, i.e., the system 
	has an evolution loop \cite{fernandez-94} of period $T$.
	\item[4.] For the cranked Hamiltonians, $H(t)=e^{-iKt}H_0e^{iKt}$,
	the cranking operator $K$ is geometrically equivalent to $H(t)$. 
	Therefore, although it is time-independent, it leads to nontrivial cyclic
	evolutions and geometric phases.
	\item[5.] Any time-independent simple harmonic oscillator admits a periodic
	dynamical invariant whose period is half the characteristic period of 
	the oscillator. We construct this invariant and the associated exact cyclic
	states, and show that they acquire nontrivial cyclic geometric phases.
	 \end{itemize}

The organization of the paper is as follows. In Section~2, we present a brief review of 
dynamical invariants and their relationship with geometric phases. In Section~3, we address 
the characterization of the geometrically equivalent quantum systems and develop a new 
approach to identify a class of exactly solvable quantum systems. In Section~4, we consider 
the quantum systems which are geometrically equivalent to a cranked Hamiltonian. In
Section~5, we apply our general results to study a complete set of nonstationary cyclic states 
of a time-independent simple harmonic oscillator. In Section~6, we present our concluding 
remarks.

\section{Dynamical Invariants and Geometric Phases}

Consider a time-dependent Hamiltonian $H(t)$ with the following
properties.
	\begin{itemize}
	\item[1.] $H(t)$ is obtained from a Hermitian parametric Hamiltonian 
$H[R]$ as $H(t)=H[R(t)]$, where $R$ stands for $(R^1,R^2,\cdots,R^d)$, 
$R^i$ are real parameters denoting the coordinates of points of a parameter 
manifold $M$, and $R(t)$ describes a smooth curve lying in $M$;
	\item[2.] The spectrum of $H[R]$ is discrete for all $R\in M$;
	\item[3.] In local patches of $M$, the eigenvalues $E_n[R]$ of $H[R]$  
are smooth functions of $R$;
	\item[4.] The curve $R(t)$ defining $H(t)$ is such that during the evolution of 
the system, i.e., for all $t\in[0,\tau]$, $E_m[R(t)]=E_n[R(t)]$ if and only if $m=n$.
In particular, the degree of degeneracy of the energy eigenvalues do not 
depend on time, and no level-crossings occur.
	\item[5.] In local patches of $M$,  there is a set of orthonormal basis 
vectors $|n,a;R\kt$ of $H[R]$ that are smooth functions of $R$.
	\end{itemize}
Here $a$ is a degeneracy label taking its values in $\{1,2,\cdots,\cun\}$ and $\cun$
denotes the degree of degeneracy of $E_n(t):=E_n[R(t)]$.

By definition, a dynamical invariant $I(t)$ of the Hamiltonian $H(t)$ is a nontrivial 
solution of the Liouville-von-Neumann equation\footnote{Here `nontrivial' means that
$I(t)$ is not a multiple of the identity operator.} \cite{lewis,lewis-riesenfeld}
	\be
	\frac{d}{dt}\,I(t)=i[I(t),H(t)]\;.
	\label{dyn-inv}
	\ee

Consider a Hermitian dynamical invariant $I(t)$ with a discrete spectrum and
let $\{|\lambda_n,a;t\kt\}$ be a complete set of orthonormal eigenvectors of
$I(t)$, where $a\in\{1,2,\cdots,d_n\}$ is a degeneracy label and $d_n$ is the
degree of degeneracy of the eigenvalue $\lambda_n$. Then one can show that
\cite{jpa-98}:
	\begin{itemize}
	\item[1.] The eigenvalues $\lambda_n$ of $I(t)$ are constant.
	\item[2.] One can express the evolution operator of the Hamiltonian $H(t)$ 
	according to
	\be
	U(t)=\sum_n \sum_{a,b=1}^{d_n} u^n_{ab}(t)|\lambda_n,a;t\kt\br
	\lambda_n,b;0|\;,
	\label{U}
	\ee
where $u^n_{ab}(t)$ are the entries of a unitary matrix $u^n(t)$ that is determined 
by the matrix Schr\"odinger equation:
	\be
	i\frac{d}{dt}\,u^n(t)=\Delta^n(t)u^n(t),~~~~u^n(0)=1\;,
	\label{u}
	\ee
with 
	\bea
	\Delta^n(t)&:=&{\cal E}^n(t)-{\cal A}^n(t)\;,
	\label{Delta}\\
	{\cal E}^n_{ab}&:=& \br\lambda_n,a;t|H|\lambda_n,b;t\kt\;,
	\label{E}\\
	{\cal A}^n_{ab}&:=&i \br\lambda_n,a;t|\frac{d}{dt}|\lambda_n,b;t\kt\;.
	\label{A}
	\eea
	\end{itemize}

For a nondegenerate eigenvalue $\lambda_n$ of the invariant, ${\cal E}^n(t)$, 
${\cal A}^n(t)$, and $\Delta^n(t)$ are scalar functions, and $u^n(t)$ is a phase factor
given by
	\bea
	u^n(t)&=&e^{i\delta_n(t)}e^{i\gamma_n(t)}\;,
	\label{u-n}\\
	\delta_n(t)&:=& -\int_0^t {\cal E}^n(s)ds=
	-\int_0^s\br\lambda;s|H(s)|\lambda_n;s\kt ds\;,
	\label{delta-n}\\
	\gamma_n(t)&:=&\int_0^t {\cal A}^n(s)ds=
	i\int_0^s\br\lambda;s|\frac{d}{ds}|\lambda_n;s\kt ds\;.
	\label{gamma-n}
	\eea
Furthermore, the state vector
	\be
	|\psi(t)\kt:=u^n(t)|\lambda_n;t\kt
	\label{psi-n}
	\ee
is an exact solution of the Schr\"odinger equation:
	\be
	i\frac{d}{dt}|\psi(t)\kt=H(t)|\psi(t)\kt\;.
	\label{sch-eq}
	\ee

For a $T$-periodic invariant, where $I(t+T)=I(t)$, the initial state vectors 
$|\psi(0)\kt$ defined by (\ref{psi-n}) undergo cyclic evolutions, and the
phase factors $e^{i\delta_n(T)}$ and $e^{i\gamma_n(T)}$ are respectively 
called the (nonadiabatic) cyclic dynamical and geometric phases \cite{abelian}. 

As shown in Ref.~\cite{jpa-98,nonabelian}, a similar analysis can be performed for the 
degenerate eigenvalues $\lambda_n$. This yields an expression for the 
non-Abelian cyclic geometric phase, namely
	\be
	\Gamma^n(T)={\cal T}e^{i\int_0^T{\cal A}^n(t)dt}\;,
	\label{Gamma}
	\ee
where ${\cal T}$ stands for the time-ordering operator. Noncyclic Abelian and non-Abelian
geometric phases can also be treated in terms of dynamical invariants 
\cite{jpa-99,nova}.

Finally, we note that in view of Eq.~(\ref{U}) any dynamical invariant satisfies
	\be
	I(t)=U(t)I(0)U^\dagger(t)\;.
	\label{i=uiu}
	\ee

\section{Geometrically Equivalent Quantum Systems}

Consider a solution $|\psi(t)\kt$ of the Schr\"odinger equation~(\ref{sch-eq}) 
for the Hamiltonian $H(t)$. Then the corresponding pure state $\Lambda(t)$ 
may be described by the projection operator $|\psi(t)\kt\br\psi(t)|$. As time
progresses, the state $\Lambda(t)$ traverses a path in the projective 
Hilbert space ${\cal P}({\cal H})$. As we mentioned in Section~1, distinct 
quantum systems may yield the same path $\Lambda(t)$ in 
${\cal P}({\cal H})$ and the same geometric phases \cite{aa}.\footnote{Note
that this is also true for noncyclic geometric phases \cite{jpa-99}.} 

Next, we suppose that $|\psi(t)\kt$ is given by Eq.~(\ref{psi-n}), i.e., it is an eigenvector of a 
dynamical invariant $I(t)$ with a nondegenerate eigenvalue $\lambda_n$.\footnote{Note 
that this is always possible. That is for any solution $|\psi(t)\kt$, one can construct a 
dynamical invariant $I(t)$ with this property.} Because, the information about the geometric 
phase is entirely included in $I(t)$, different Hamiltonians admitting $I(t)$ as a dynamical
invariant would lead to the same set of geometric phases (for the evolution of the eigenstates 
of $I(0)$.) This observation provides the means for a characterization of the geometrically 
equivalent quantum systems in terms of the symmetries of dynamical invariants. 

Specifically, consider another Hamiltonian $\tilde H(t)$ with the same spectral 
properties as $H(t)$, and suppose that $I(t)$ is an arbitrary Hermitian
dynamical invariant of both $H(t)$ and $\tilde H(t)$, i.e., $I(t)$ satisfies Eqs.~(\ref{dyn-inv}) and
	\be
	\frac{d}{dt}\,I(t)=i[I(t),\tilde H(t)]\;.
	\label{dyn-inv-tilde}
	\ee
Then $H(t)$ and $\tilde H(t)$ are geometrically equivalent. Next, introduce 
$X(t):=\tilde H(t)-H(t)$. In view of Eqs.~(\ref{dyn-inv}) and (\ref{dyn-inv-tilde}),
	\be
	[I(t),X(t)]=0\;.
	\label{IX}
	\ee
In other words, the two Hamiltonians $H(t)$ and $\tilde H(t)$  are related by a
symmetry generator $X(t)$ of  the invariant $I(t)$ according to
	\be
	\tilde H(t)=H(t)+X(t)\;.
	\label{X}
	\ee

Furthermore, it is not difficult to show that the evolution operator $\tilde U(t)$
of $\tilde H(t)$ may be written in the form
	\be
	\tilde U(t)=\sum_n \sum_{a,b=1}^{d_n} \tilde u^n_{ab}(t)
	|\lambda_n,a;t\kt\br	\lambda_n,b;0|\;,
	\label{U-tilde}
	\ee
where $\tilde u^n(t)$ is defined by
	\bea
	i\frac{d}{dt}\,\tilde u^n(t)&=&\tilde\Delta^n(t)u^n(t)\,,~~~~\tilde u(0)=1\;,
	\label{u-tilde}\\
	\tilde\Delta^n(t)&:=&\tilde{\cal E}^n(t)-{\cal A}^n(t)\;,
	\label{Delta-tilde}\\
	\tilde{\cal E}^n_{ab}&:=& \br\lambda_n,a;t|\tilde H|\lambda_n,b;t\kt\;.
	\label{E-tilde}
	\eea
Note that the transformation $H(t)\to\tilde H(t)$ leaves the matrices ${\cal A}^n(t)$ and 
consequently the geometric phases invariant.

Eqs.~(\ref{i=uiu}), (\ref{X}) and (\ref{U-tilde}) suggest that given a 
time-dependent Hamiltonian $H(t)$ whose Schr\"odinger equation is
exactly solvable, one can obtain the exact solution of the Schr\"odinger 
equation for all the geometrically equivalent Hamiltonians $\tilde H(t)$.
For example, let $X(t)$ be a polynomial in $I(t)$, i.e.,
	\be
	X(t)= \sum_{i=1}^p f_i(t)[I(t)]^i\;,
	\label{X=}
	\ee
where $f_i$ are arbitrary smooth real-valued functions of $t$. Then $X(t)$ 
commutes with $I(t)$ and one obtains a class of exactly solvable 
time-dependent Hamiltonians given by
	\be
	\tilde H(t)=H(t)+\sum_{i=1}^n f_i(t)[I(t)]^i\;.
	\label{exact}
	\ee

Next, consider an arbitrary symmetry generator $X(t)$ of $I(t)$, and let
	\be
	Y(t):=U^\dagger(t)X(t)U(t)\;,
	\label{Y1}
	\ee
where $U(t)$ is the evolution operator for the Hamiltonian $H(t)$, i.e., the solution 
of 
	\be
	i\frac{d}{dt}\,U(t)=H(t)U(t)\;,~~~~U(0)=1\,.
	\label{U-def}
	\ee
In view of Eqs.~(\ref{i=uiu}), (\ref{Y1}), (\ref{X}), and (\ref{U-def}), we have
	\bea
	[Y(t),I(0)]&=&0\;,
	\label{y-i}\\
	\tilde H(t)&=&U(t)Y(t)U^\dagger(t)-i\dot U(t)U^\dagger(t)\;.
	\label{H-trans}
	\eea
These equations indicate that the Hamiltonian $\tilde H(t)$ is related to a symmetry
generator of the initial invariant $I(0)$ through a (canonical) unitary transformation
of the Hilbert space \cite{jmp-97,nova}:
	\be
	|\phi(t)\kt\to|\tilde\psi(t)\kt:=U(t)|\phi(t)\kt\;,
	\label{phi-trans}
	\ee
where $|\phi(t)\kt$ is a solution of the Schr\"odinger equation with $Y(t)$ playing the
role of the Hamiltonian, i.e.,
	\be
	|\phi(t)\kt=V(t)|\phi(0)\kt\;,~~~~V(t)={\cal T}\,e^{-i\int_0^t Y(s)ds}\;.
	\label{V=}
	\ee
In view of this observation, the evolution operator $\tilde U(t)$ of $\tilde H(t)$ is related
to the evolution operator $V(t)$ of $Y(t)$ according to
	\be
	\tilde U(t)=U(t)V(t)\;.
	\label{u=uv}
	\ee
In particular, if $Y(t)$ is a constant operator, 
	\be
	\tilde U(t)=U(t) e^{-itY}\;.
	\label{u=uv2}
	\ee
Furthermore, if $X(t)$ is given by Eq.~(\ref{X=}), then in light of Eqs.~(\ref{i=uiu}) and
(\ref{Y1}),
	\bea
	Y(t)&=&\sum_{i=1}^pf_i(t)[I(0)]^i\;,
	\label{Y2}\\
	\tilde U(t)&=& U(t)e^{-i\int_0^t Y(s)ds}=U(t)e^{-i\sum_{i=1}^p F_i(t)[I(0)]^i}\;,
	\label{U=U-sum}
	\eea
where $F_i(t):=\int_0^tf_i(s)ds$.

Eq.~(\ref{u=uv}) provides the desired relationship between the evolution operators of
geometrically equivalent Hamiltonians.

In the remainder of this section, we address the question of characterizing the
class of all the Hamiltonians that admit a given dynamical invariant $I(t)$ with 
the following properties \cite{jpa-99,nova}.
	\begin{itemize}
	\item[1.] $I(t)$ is obtained from a Hermitian parametric invariant 
$I[\bar R]$ as $I(t)=I[\bar R(t)]$, where $\bar R$ stands for $(\bar R^1,\bar R^2,
\cdots,\bar R^{\bar d})$, $\bar R^i$ are real parameters denoting the coordinates
of points of a parameter manifold $\bar M$, and $\bar R(t)$ describes a smooth 
curve lying in $\bar M$;
	\item[2.] The spectrum of $I[\bar R]$ is discrete for all $\bar R\in\bar M$;
	\item[3.] In local patches of $\bar M$,  there is a set of orthonormal basis 
vectors $|\lambda_n,a;\bar R\kt$ of $I[\bar R]$ that are smooth (single-valued) functions of 
	$\bar R$.
	\end{itemize}

Clearly one can express the parametric invariant $I[\bar R]$ in the form
	\be
	I[\bar R]=\bar W[\bar R] I_0\bar W^\dagger[\bar R]\;,
	\label{i=wiw}
	\ee
where $\bar W[\bar R]$ is defined by the condition
	\be
	|\lambda_n,a;\bar R\kt=\bar W[\bar R]|\lambda_n,a;\bar R(0)\kt\;,
	\label{W}
	\ee
and $I_0:=I[\bar R(0)]$. Now, it is not difficult to check that for any curve $\bar R(t)$
in $\bar M$, the invariant $I(t):=I[\bar R(t)]$ satisfies the Liouville-von-Neumann 
equation~(\ref{dyn-inv}) for the Hamiltonian
	\be
	H_*(t):=i\dot{\bar W}[\bar R(t)]\bar W^\dagger[\bar R(t)]\;,
	\label{H-star}
	\ee
where a dot denotes a time-derivative. Moreover, any other Hamiltonian that admits the
invariant $I(t)$ is of the form
	\be
	H(t)=H_*(t)+X(t)\;,
	\label{characterize}
	\ee
where $X(t):=X[\bar R(t)]$ and $X[\bar R]$ is any Hermitian operator that commutes with
$I[\bar R]$. 

This completes the characterization of the geometrically equivalent quantum systems. 

We conclude this section with the following remarks. 

\begin{itemize}
\item[I.] Substituting Eqs.~(\ref{W}) and (\ref{H-star}) in Eqs.~(\ref{E}) and (\ref{A}),
we find that for the Hamiltonian $H_*(t)$,
	\[{\cal E}_{ab}^n(t)={\cal A}_{ab}^n(t)=i\br\lambda_n,a;\bar R(0)|
	\bar W^\dagger[\bar R(t)]\dot{\bar W}[\bar R(t)]|\lambda_n,b;\bar R(0)\kt\;.\]
Therefore, $\Delta^n(t)=0$ and 
	\be
	u^n(t)=1.
	\label{u=1}
	\ee
In particular, the evolution operator for $H_*(t)$ is given by
	\be
	U_*(t)=\sum_n\sum_{a=1}^{d_n}|\lambda_n,a;t\kt\br\lambda_n,a;0|=W[\bar R(t)]\;.
	\label{U-H-star}
	\ee
Moreover, for a nondegenerate eigenvalue $\lambda_n$, the geometric and dynamical 
phases cancel each other.
\item[II.] Let $Z[\bar R]$ be any unitary operator commuting with $I_0$. Then the 
transformation 
	\be
	\bar W[\bar R]\to\bar W'[\bar R]:=\bar W[\bar R] Z[\bar R]
	\label{gauge}
	\ee
leaves the form of $I[\bar R]$ unchanged. In other words, the operators $\bar W[\bar R]$
are subject to `gauge transformations'~(\ref{gauge}). These transformations are essentially
the transformations of the basis vectors of the degeneracy subspaces of the invariant,
	\be
	|\lambda_n,a;\bar R\kt \to |\lambda_n,a;\bar R\kt':=
	\bar W'[\bar R]|\lambda_n,a;\bar R(0)\kt.
	\label{basis-trans}
	\ee
The physical quantities are invariant under these transformations. Therefore, they may be
calculated after making a choice for the gauge: $\bar W[\bar R]$. The choice of 
$\bar W[\bar R]$ is only restricted in the sense that it must satisfy Eq.~(\ref{i=wiw})
and be a single-valued (differentiable) function of $\bar R$. Note also that the 
 Hamiltonian~(\ref{H-star}) and its evolution operator are not invariant under the gauge 
transformations~(\ref{gauge}); they transform according to
	\be
	H_*(t)\to H'_*(t)=H_*(t)+i\bar W[\bar R(T)]\dot Z[\bar R(t)]Z[\bar R(t)]^\dagger
	W[\bar R(t)]^\dagger,~~~~U_*(t)\to U'_*(t)=U_*(t)Z(t),
	\label{H-U}
	\ee
respectively. However, it is not difficult to show that the relation (\ref{u=1}) 
survives these transformations. 
\item[III.] Suppose that the invariant $I(t)$ is periodic. Then one may choose a gauge in 
which 
	\be
	\bar W[\bar R(T)]=\bar W[\bar R(0)]=1.
	\label{periodic}
	\ee
This may be realized by requiring that the curve $\bar R(t)$ associated with the invariant is 
closed, i.e., there is $T\in\R^+$ such that $\bar R(T)=\bar R(0)$. This in turn implies that 
$H_*(T)=H_*(0)$. Thus the class of geometrically equivalent quantum systems determined 
by the $T$-periodic invariant includes a system with a $T$-periodic Hamiltonian $H_*(t)$. 
Furthermore, according to Eqs.~(\ref{periodic}) and (\ref{U-H-star}), $U_*(T)=1$. This 
means that quantum system described by the $T$-periodic Hamiltonian $H_*(t)$ 
has an evolution loop of period $T$, \cite{fernandez-94}. Note, however, that according to
Eqs.~(\ref{H-U}), in an arbitrary gauge, $H_*(t)$ is not $T$-periodic. Yet its evolution
operator satisfies $[U(T),I(0)]=0$. This is actually a necessary and sufficient condition for
the vectors $|\lambda_n,a;\bar R(0)\kt$ to perform cyclic evolutions, \cite{jpa-98}.
\end{itemize}

\section{Quantum Systems that Are Geometrically Equivalent to a Cranked System}

By definition, a cranked Hamiltonian has the form
	\be
	H(t)=e^{-iK t}H_0e^{iKt} \;,
	\label{c1}
	\ee
where $K$ and $H_0$ are constant Hermitian operators. It is not difficult to show that
	\be
	I(t):=H(t)-K=e^{-iK t}(H_0-K)e^{iKt}\;,
	\label{c2}
	\ee
is a dynamical invariant for the cranked Hamiltonian~(\ref{c1}). Furthermore, one can
perform a time-dependent unitary transformation to map the system to a canonically
equivalent system with a constant Hamiltonian \cite{jmp-97,jpa-98a}. This method yields
the following expression for the evolution operator of the cranked 
Hamiltonian~(\ref{c1}):\footnote{For details see \cite{nova}.}
	\be
	U(t)=e^{-iKt}e^{-i(H_0-K)t}\;.
	\label{c3}
	\ee

Next, we consider the class of Hamiltonians that are geometrically equivalent to a cranked
Hamiltonian. According to Eqs.~(\ref{X}) and (\ref{c2}), these Hamiltonians have the following general form:
	\be
	\tilde H(t)=e^{-iK t}[H_0+Y(t)]e^{iKt}\;,
	\label{X1}
	\ee
where $Y(t)$ is any Hermitian operator commuting with
	\be
	I_0=I(0)=H_0-K\,.
	\label{c4.5}
	\ee

Now, let $\tilde Y(t)$ be any operator commuting with $I(0)$ and set 
$Y(t)=-I(0)+\tilde Y(t)=K-H_0+\tilde Y(t)$. Then Eq.~(\ref{X1}) yields
	\be
	\tilde H(t)=	e^{-iK t}[K+\tilde Y(t)]e^{iKt}\;.
	\label{X2}
	\ee
In particular, setting $\tilde Y(t)=0$, we find that the cranking operator $K$ is also 
geometrically equivalent to the cranked Hamiltonian~(\ref{c1}). Note that although $K$ is 
time-independent, it admits a time-dependent invariant, namely~(\ref{c2}). This in turn 
implies that as the eigenstates of this invariant evolve in time (according to the 
Schr\"odinger equation defined by $K$), they develop nontrivial (cyclic and noncyclic) 
geometric phases. Therefore, the above construction provides a simple example of a
time-independent Hamiltonian leading to nontrivial geometric phases 
\cite{moore-90,bohm-qm,fernandez-94}.

Furthermore, we can use an argument similar to the one leading to Eq.~(\ref{u=uv})
to express the evolution operator $\tilde U(t)$ of $\tilde H(t)$ in terms of the evolution 
operator $e^{-iKt}$ of $K$. This yields
	\be
	\tilde U(t)=e^{-iKt}{\cal T}\, e^{-i\int_0^t \tilde Y(s)ds}\;.
	\label{tilde-U2}
	\ee

Next, we compute geometric phases associated with the eigenstates of the 
invariant~(\ref{c2}). This requires the computation of the eigenvectors $|\lambda_n,a;t\kt$ 
or alternatively the unitary operators $\bar W[\bar R(t)]$. We wish to emphasize that one 
must refrain from identifying the operator $\bar W[\bar R(t)]$ of Eq.~(\ref{i=wiw}) with 
$e^{-iKt}$. In order to obtain $\bar W[\bar R(t)]$, one must first determine the parameter 
space of the invariant and the single-valued functions $\bar W[\bar R]$. In general, the 
choice $\bar W[\bar R(t)]=e^{-iKt}$ violates the requirement that $\bar W[\bar R]$ must be
single-valued. In view of Eqs.~(\ref{i=wiw}) and (\ref{c2}), we can however write
	\be
	\bar W[\bar R(t)]=e^{-iKt}Z(t)\;,
	\label{W=KZ}
	\ee
where $Z(t)$ is a unitary operator commuting with $I(0)$. 

Inserting Eq.~(\ref{W=KZ}) in Eqs.~(\ref{H-star}) and (\ref{W}), we obtain 
	\bea
	H_*(t)&=&K+ ie^{-iKt}\dot Z(t)Z^\dagger(t)e^{iKt}\;.
	\label{c4-new}\\
	|\lambda_n,a;t\kt&=&e^{-iKt}Z(t)|\lambda_n,a;0\kt.
	\label{c5}
	\eea
Because $Z(t)$ and $\tilde Y(t)$ commute with $I(0)$, there exist scalar complex-valued 
functions $z_n^{ab}$ and $\tilde y_n^{ab}$ satisfying
	\be
	Z(t)|\lambda_n,a;0\kt=\sum_{b=1}^{d_n}z_n^{ba}(t)|\lambda_n,b;0\kt\;,~~~~
	\tilde Y(t)|\lambda_n,a;0\kt=\sum_{b=1}^{d_n}\tilde y_n^{ba}(t)|\lambda_n,b;0\kt\,.
	\label{Z1}
	\ee

Now, we are in a position to compute the matrices ${\cal E}^n$,  ${\cal A}^n$, and
$\Delta^n$ for the Hamiltonian~(\ref{X2}). The result is 
	\bea
	{\cal E}^n(t)&=& Z_n^\dagger(t)K_nZ_n(t) + Z_n^\dagger(t)\tilde Y_n(t)Z_n(t)\;,
	\label{cc6}\\
	{\cal A}^n(t)&=&Z_n^\dagger(t)K_nZ_n(t)+ iZ_n^\dagger(t)\dot Z_n(t)\;,
	\label{cc7}\\
	\Delta^n(t)&=&Z_n^\dagger(t)\tilde Y_n(t)Z_n(t)-iZ_n^\dagger(t)\dot Z_n(t)\;,
	\label{cc8}
	\eea
where $Z_n,K_n,$ and $\tilde Y_n$ are $d_n\times d_n$ matrices with entries $z_n^{ab},
\br\lambda_n,a;0|K|\lambda_n,b;0\kt$, and $\tilde y_n^{ab}$, respectively. 
	
Note that $\Delta^n$ is related to $\tilde Y_n$ by a time-dependent (canonical) unitary
transformation. This, in particular, implies
	\be
	u^n(t)=Z_n^\dagger(t){\cal T}\,e^{-i\int_0^t\tilde Y_n(s)ds}\;.
	\label{cc9}
	\ee
Substituting Eqs.~(\ref{c5}) and (\ref{cc9}) in Eq.~(\ref{U}), we recover 
Eq.~(\ref{tilde-U2}).

For a nondegenerate eigenvalue $\lambda_n$, we have $Z_n(t)=e^{-i\zeta_n(t)}$ where
$\zeta_n(t)\in\R$ and
	\[{\cal E}^n(t)=K_n+\tilde Y_n(t)\;,~~~~{\cal A}^n(t)=K_n+ \zeta_n(t)\;,
	~~~~u^n(t)=e^{i\zeta_n(t)} e^{-i\int_0^t\tilde Y_n(s)ds}\;.\]
In particular, if $I(t)$ is $T$-periodic, the cyclic geometric and dynamical phase angles are
given by
	\[\gamma_n(T)=K_n T + \zeta_n(T)\,,~~~~\delta_n(T)=-K_n T-\int_0^T
	\tilde Y_n(t)dt\;.\]

\section{Cyclic States and Geometric Phases for a Time-Independent Simple Harmonic 
Oscillator}

Consider the cranked Hamiltonian~(\ref{c1}) defined by the initial Hamiltonian
	\be
	H_0:=\frac{p^2}{2M}+\frac{M\Omega^2}{2}\, x^2\;,
	\label{h1}
	\ee
and the cranking operator
	\be
	K:=\frac{p^2}{2m}+\frac{m\omega^2}{2}\, x^2\;,
	\label{h2}
	\ee
where $M,\Omega,m,\omega$ are positive real numbers satisfying 
	\be
	m>M~~~{\rm and}~~~ M\Omega^2>m\omega^2\;.
	\label{h3}
	\ee

It is not difficult to compute the cranked Hamiltonian~(\ref{c1}). First we use the 
Backer-Campbell-Hausdorff formula to establish the identities
	\bea
	e^{-iKt}\,x\,e^{iKt}&=&\cos(\omega t)\,x-(m\omega)^{-1}\sin(\omega t)p\;,
	\label{h4}\\
	e^{-iKt}\,p\,e^{iKt}&=&
	m\omega\sin(\omega t) x+\cos(\omega t)\,p\;.
	\label{h5}
	\eea
Substituting Eqs.~(\ref{h1}) and (\ref{h2}) in Eq.~(\ref{c1}) and making use of 
Eqs.~(\ref{h4}) and (\ref{h5}), we have
	\be
	H(t)= \frac{1}{2}\left\{[a+b\cos(2\omega t)]\,p^2+
	[c\sin(2\omega t)](xp+px)+[d+e\cos(2\omega t)]x^2\right\}\;,
	\label{h6}
	\ee
where
	\bea
	&& 
	a:=\frac{1+\nu^2}{2M}\,,~~~~
	b:=\frac{1-\nu^2}{2M}\,,~~~~
	c:=\frac{m\omega(1-\nu^2)}{2M}\,,\nn\\
	&&
	d:=\frac{(m\omega)^2(1+\nu^2)}{2M}\,,~~~~
	e:=-\frac{(m\omega)^2(1-\nu^2)}{2M}\,,~~~~
	\nu:=\frac{M\Omega}{m\omega}\,.\nn
	\eea

As seen from Eq.~(\ref{h6}), $H(t)$ is the Hamiltonian for a periodic time-dependent 
generalized harmonic oscillator of period $T=\tau/2$ where $\tau:=2\pi/\omega$ is the 
characteristic period of the harmonic oscillator described by the Hamiltonian~(\ref{h2}). 
According to Eq.~(\ref{c2}), the Hamiltonian~(\ref{h6}) admits a dynamical invariant of the 
form 
	\be
	I(t)=H(t)-K=\frac{1}{2}\left\{[a-m^{-1}+b\cos(2\omega t )]\,p^2+
	[c\sin(2\omega t)](xp+px)+[d-m\omega^2 +e\cos(2\omega t)]x^2\right\}\;,
	\label{h7}
	\ee
which is also $T$-period. Furthermore, in view of the results of the preceding section, 
$I(t)$ is also a dynamical invariant for the Hamiltonian $K$. In other words, we have 
constructed a nontrivial $T$-periodic dynamical invariant~(\ref{h7}) for a time-independent
simple harmonic oscillator with arbitrary mass $m$ and frequency $\omega$.

By construction, the eigenstates of the initial invariant $I(0)=H_0-K$ perform exact 
cyclic evolutions of period $T=\tau/2$. Note that by virtue of conditions~(\ref{h3}), we 
can identify $I(0)$ with the Hamiltonian of a simple Harmonic oscillator of mass 
$\tilde m:=(M^{-1}-m^{-1})^{-1}$ and frequency 
$\tilde\omega:=\sqrt{(M^{-1}-m^{-1})(M\Omega^2-m\omega^2)}$. Therefore, we can 
easily obtain the expression for its eigenvectors \cite{sakurai}:
	\be
	|\lambda_n;0\kt=(n!)^{-1/2}\tilde a^{\dagger n}|0\kt,
	\label{h8}
	\ee
where
	\[\tilde a:=\sqrt{\frac{\tilde m\tilde\omega}{2}}
	\left(x+\frac{ip}{\tilde m\tilde\omega}\right)\;,~~~~
	\br x|0\kt=\left(\frac{\tilde m\tilde\omega}{\pi}\right)^{1/4}e^{-\tilde m\tilde\omega
	x^2/2}\;.\]

Note that $|\lambda_n;0\kt$ are not eigenvectors of the Hamiltonian $K$. Because $K$ 
does not depend on time, the corresponding evolution operator is given by ${\cal U}(t)=
e^{-iKt}$. Therefore, the initial eigenvectors $|\lambda_n;0\kt$ evolve according to
	\be
	|\psi_n(t)\kt=e^{-iKt}|\lambda_n;0\kt\;.
	\label{h9}
	\ee
Because $ |\lambda_n;0\kt $ are eigenvectors of $I(0)$, the corresponding pure states
perform cyclic evolutions of period $T$;
	\[|\lambda_n;T\kt\br\lambda_n;T| = |\lambda_n;0\kt\br\lambda_n;0|.\]
Hence $|\lambda_n;0\kt$ form a complete orthonormal set of basis vectors of the Hilbert 
space that undergo (nonstationary) cyclic evolutions of period $T=\tau/2=\pi/\omega$. 

In order to compute the cyclic geometric and dynamical phases associated with these cyclic
states, we need to determine a complete set of orthonormal basis vectors $|\lambda_n;t\kt$ 
of $I(t)$. We first note that the harmonic oscillator~(\ref{h2}) does not have an evolution 
loop of period $T$. The period of the evolution loops of a time-independent simple 
harmonic oscillator is an integer multiple $n\tau$ of its characteristic period $\tau$, where 
$n$ is a positive integer and $U(n\tau)=(-1)^n$, \cite{benedict-schleich}. In particular, 
$e^{-iKT}=e^{-iK\tau/2}$ is not a multiple of the identity operator, and $e^{-itK}$ is not 
$T$-periodic. This is an indication that the operator $Z(t)$ of Eq.~(\ref{W=KZ}) is 
different from the identity operator. In order to determine $Z(t)$ or alternatively the 
single-valued unitary operators $\bar W[\bar R]$, we need to investigate the parameter space 
of the corresponding parametric invariant.

We first introduce
	\be
	I[\bar R]=\bar b\sum_{i=1}^n \bar R^iK_i,
	\label{su0}
	\ee
where $\bar b$ is a constant, $\bar R=(\bar R^1,\bar R^2,\bar R^3)$ are parameters of the invariant, and
	\bea
	K_1&:=&\frac{1}{4}(X^2-P^2)\;,
	\label{su1}\\
	K_2&:=&-\frac{1}{4}(XP+PX)=-\frac{1}{4}(xp+px)\;,
	\label{su2}\\
	K_3&:=&\frac{1}{4}(X^2+P^2)\;,
	\label{su3}\\
	X&:=&\sqrt{\tilde m\tilde\omega}\,x,~~~~
	P\,:=\,\frac{p}{\sqrt{\tilde m\tilde \omega}}.
	\label{su4}
	\eea
Note that $K_i$ are the generators of the group $SU(1,1)$, i.e., they satisfy
	\be
	[K_1,K_2]=-iK_3,~~~~
	[K_2,K_3]=iK_1,~~~~
	[K_3,K_1]=-iK_2.
	\label{su5}
	\ee
The parameter space of the invariant~(\ref{su0}) is the unit hyperboloid:\footnote{For 
a detailed treatment of the parameterization of the quadratic invariants of harmonic
oscillators and related issues, see Ref.~\cite{nova}.}
	\[\left\{\bar R\in\R^3~|~-(\bar R^1)^2-(\bar R^2)^2+(\bar R^3)^2=1,~
	\bar R^3>0\right\}.\]
It is convenient to express $I[\bar R]$ in the hyperbolic coordinates 
	\be
	\bar\theta=\cosh^{-1}(\bar R^3)\in\R\,,
	~~~~\bar\varphi=\tan^{-1}(\bar R^2/\bar R^1)\in[0,2\pi).
	\label{h30}
	\ee
This yields
	\be
	I[\bar R]=I[\bar\theta,\bar\varphi]=b ( \sinh\bar\theta\cos\bar\varphi K_1+
	\sinh\bar\theta\sin\bar\varphi K_2+\cosh\bar\theta K_3).
	\label{hyper}
	\ee

Next, we introduce 
	\be
	\varphi(t):=2\omega t.
	\label{phi}
	\ee
and write the invariant (\ref{h7}) in the form $I[\bar R(t)]$, where
	\bea
	\bar b&=&2\tilde\omega\,,~~~~
	\bar R^1(t)=\frac{1}{2} \left(\tilde m b-\frac{e}{\tilde m\tilde\omega^2}\right)
	[1-\cos\varphi(t)]\,,
	\label{cu1}\\
	\bar R^2(t)&=&-\left(\frac{c}{\tilde\omega}\right)\sin\varphi(t)\;,~~~~
	\bar R^3(t)=1-\frac{1}{2} \left(\tilde m b+\frac{e}{\tilde m\tilde\omega^2}\right)
	[1-\cos\varphi(t)]\,,
	\label{cu2}
	\eea
In view of Eqs.~(\ref{h30}), (\ref{cu1}) and (\ref{cu2}), we have
	\be
	\cosh\bar\theta(t)=1+\zeta[1-\cos\varphi(t)]\,,~~~~
	\tan\bar\varphi(t)=\frac{\xi\sin\varphi(t)}{1-\cos\varphi(t)}\;,
	\label{h31}
	\ee
where 
	\bea
	\zeta&:=&-\frac{1}{2}\left(\tilde m b+\frac{e}{\tilde m\tilde\omega^2}\right)\,=\,
	-\frac{(1-\nu^2)(1-\mu^2)}{4(1-\frac{M}{m})}\;,\nn\\
	\xi&:=&-\frac{2c}{\tilde m\tilde\omega b-\frac{e}{\tilde m\tilde\omega}}\,=\,
	-\frac{2\mu}{1+\mu^2}\;,\nn\\
	\mu&:=&\frac{m\omega}{\tilde m\tilde\omega}\;.\nn
	\eea

Now, following the same method used in Refs.~\cite{nakagawa,bohm-qm,nova} to compute 
the eigenvectors of the dipole Hamiltonian and the generalized harmonic oscillator, we use 
the commutation relations (\ref{su5}) to express the invariant~(\ref{hyper})  in the form 
(\ref{i=wiw}) with
	\be
	\bar W[\bar R] =e^{-i\bar\varphi K_3} e^{-i\bar\theta K_2}e^{i\bar\varphi K_3}\;.
	\label{w=eeee}
	\ee
One can easily check the validity of Eq.~(\ref{i=wiw}) for this choice of $\bar W[\bar R]$
by noting that 
	\[\bar W[\bar R] K_3\bar W[\bar R]^\dagger=
	\sinh\bar\theta\cos\bar\varphi K_1+\sinh\bar\theta\sin\bar\varphi K_2+
	\cosh\bar\theta K_3.\]
This equation follows from the Backer-Campbell-Hausdorff formula and Eqs.~(\ref{su5}).

Having obtained the unitary operator $\bar W[\bar R]$, we may compute 
$|\lambda_n;t\kt=\bar W[\bar R(t)]|\lambda_n;0\kt$ and the corresponding 
phase angles $\delta_n(t)$ and $\gamma_n(t)$. 

First, we note that by construction
	\be
	|\psi_n(t)\kt=e^{i\delta_n(t)}e^{i\gamma_n(t)}|\lambda_n;t\kt\;.
	\label{h33}
	\ee
This implies that
	\be
	{\cal E}^n(t)=\br\lambda_n;t|K|\lambda_n;t\kt=\br\psi_n(t)|K|\psi_n(t)=
	\br\lambda_n;0|K|\lambda_n;0\kt\;,
	\label{h34}
	\ee
where we have also made use of Eq.~(\ref{h9}). We can easily compute the write hand side 
of Eq.~(\ref{h34}) using the well-known identities \cite{messiah}:
	\[\br\lambda_n;0|x^2|\lambda_n;0\kt=(\tilde m\tilde\omega)^{-1}(n+\frac{1}{2})\,
	~~~~	\br\lambda_n;0|p^2|\lambda_n;0\kt=\tilde m\tilde\omega (n+\frac{1}{2})\,.\]
Substituting these equations in (\ref{h34}) and using Eq.~(\ref{delta-n}), we obtain
	\be
	\delta_n(t)=\delta_0(t)(2n+1)\,,~~~~~\delta_0(t)=-\frac{1}{4}\,
	(\mu+\mu^{-1})\omega t\,.
	\label{delta-sho}
	\ee
In particular, the dynamical phase angle is given by
	\[ \delta_n(T)=-\frac{\pi}{4}\,(\mu+ \mu^{-1} )(2n+1).\]

Next, we compute 
	\bea
	{\cal A}^n(t)&=&i\br\lambda_n;t|\frac{d}{dt}|\lambda_n;t\kt=
	i\br\lambda_n;0|\bar W[\bar R(t)]^\dagger\dot{\bar W}[\bar R(t)] |\lambda_n;0\kt\nn\\
	&=&\frac{1}{4}\, (2n+1) [\cosh\bar\theta(t)-1]\dot{\bar\varphi}(t).
	\label{h35}
	\eea
In the derivation of Eq.~(\ref{h35}), we have employed the identities:
	\bea
	\bar W[\bar R(t)]^\dagger\dot{\bar W}[\bar R(t)]&=&
	i[\sinh\bar\theta(\cos\bar\varphi~K_1+\sin\bar\varphi~K_2)
	(1-\cosh\bar\theta)K_3]\dot{\bar\varphi}+\nn\\
	&&i(\sin\bar\varphi~K_1-\cos\bar\varphi~K_2)\dot{\bar\theta}\,,\nn\\
	\br\lambda_n;0|K_1|\lambda_n;0\kt&=&\br\lambda_n;0|K_2|\lambda_n;0\kt=0\;,\nn\\
	\br\lambda_n;0|K_3|\lambda_n;0\kt&=&\frac{1}{4}\,(2n+1)\,.\nn
	\eea
Inserting Eq.~(\ref{h35}) in Eq.~(\ref{gamma-n}) and making use of Eq.~(\ref{phi}), we
find
	\be
	\gamma_n(t)=\frac{1}{4}\,(2n+1) \int_{0}^{t}
	[\cosh\bar\theta(t)-1] \dot{\bar\varphi}(t)dt=\frac{1}{4}\,(2n+1) 
	\int_{0}^{\varphi(t)}[\cosh\bar\theta(\varphi)-1]
	\frac{d\bar\varphi(\varphi)}{d\varphi}\, d\varphi.
	\label{h36}
	\ee
Now, we use the second equation in (\ref{h31}) to calculate
	\[\frac{d\bar\varphi(\varphi)}{d\varphi}=-\frac{\xi}{(\xi^2+1)+	(\xi^2-1)\cos\varphi}\;.\]
Substituting this equation in (\ref{h36}) and performing the integral, we finally obtain 
	\bea
	\gamma_n(t)&=&(2n+1)\gamma_0(t)\,,~~~~~
	\gamma_0(t)=\frac{\zeta\xi\sigma(t)}{4(1-\xi^2)}=
	\frac{\mu(1+\mu^2)(1-\nu^2)\sigma(t)}{8(1-\frac{M}{m})(1-\mu^2)}\;,\nn\\
	\sigma(t)&:=&-\varphi(t)+2|\xi|\tan^{-1}\left[\frac{\tan\frac{\varphi(t)}{2}}{|\xi|}\right]=
	-2\omega t+2|\xi|\tan^{-1}\left[\frac{\tan(\omega t)}{|\xi|}\right]\,.\nn
	\eea
In particular, the cyclic geometric phase angle associated with the initial state vector
$|\lambda_n;0\kt$ has the form
	\be
	\gamma_n(T)=\frac{\pi\zeta\xi(2n+1)}{2(\xi^2-1)}=
	\frac{\pi\mu(1+\mu^2)(1-\nu^2)(2n+1)}{4(1-\frac{M}{m})(\mu^2-1)}\;.
	\label{gp-sho}
	\ee

Next, we return to the class of geometrically equivalent quantum systems that include the
simple harmonic oscillator Hamiltonian~(\ref{h2}). The Hamiltonian for such a system is
given by Eq.~(\ref{X2}). If we set $\tilde Y(t)=f(t)[H_0-K]$ in this equation, where $f$ 
is an arbitrary smooth positive real-valued function of time, we find a class of time-dependent 
generalized harmonic oscillator Hamiltonians of the form
	\be
	H_{\rm GHO}(t)=f(t)H(t)+[1-f(t)]K
	\label{gho}
	\ee
with $H(t)$ given by Eq.~(\ref{h6}). Note that the Hamiltonian~$H(t)$ involves four
parameters, namely $m,M,\omega$ and $\Omega$, that must obey conditions~(\ref{h3}).
Therefore, Eq.~(\ref{gho}) determines a five-parameter family of time-dependent
generalized harmonic oscillators, where four of the parameter are real numbers and the fifth 
parameter is a function $f(t)$. In view of Eq.~(\ref{tilde-U2}), the evolution operator for 
the Hamiltonian~(\ref{gho}) is given by
	\[\tilde U(t)=e^{-iKt}e^{-iF(t)(H_0-K)]},\]
where $F(t):=\int_0^tf(s)ds$. Note that, in general, $H_{\rm GHO}(t)$ is not $T$-periodic. 
Yet it admits a $T$-periodic invariant, namely~(\ref{h7}).

\section{Conclusion}

We have investigated the quantum systems that give rise to the same set of geometric phases 
for a complete set of initial state vectors. We termed these systems geometrically equivalent. 
We argued that these systems admit a common dynamical invariant and used this observation 
to yield a complete characterization of these systems. Furthermore, we showed how the 
evolution operators of the geometrically equivalent systems are related. In particular, 
we investigated the class of cranked Hamiltonians and applied our general results to simple 
harmonic oscillators.

We addressed the characterization problem for the systems that are geometrically
equivalent to a time-independent simple harmonic oscillator. Our solution showed that 
this simple system admits periodic dynamical invariants. We used such an invariant to
construct a complete orthonormal set of initial state vectors that undergo nonstationary
cyclic evolutions. These states involve nontrivial geometric and dynamical phases.

The invariant~(\ref{h7}) that we constructed for the simple harmonic oscillator~(\ref{h2})
may be put in the form \cite{lewis-riesenfeld}
	\[I(t)=\frac{1}{2}\left[(\rho p-m\dot\rho x)^2+\rho^{-2}x^2\right]\,,\]
where 
	\be
	\rho:=\sqrt{\tilde m^{-1}-b[1-\cos(2\omega t)]}
	\label{rho}
	\ee
satisfies the Ermakov equation \cite{ermakov}
	\be
	\ddot\rho+\omega^2\rho=\frac{\eta}{\rho^3}\;,
	\label{ermakov}
	\ee
with $\eta:=\tilde m^{-1}(\tilde m^{-1}-2b)\omega^2$. This is indeed to be expected 
for the general solution of Eq.~(\ref{ermakov}) may be written \cite{pinney} in the form
$\rho=\sqrt{c_1x_1^2(t)+c_2x_2^2(t)}$ where $c_1$ and $c_2$ are constants and $x_1$ and 
$x_2$ are two linearly independent solutions of the classical equation of motion 
$\ddot x+\omega^2 x=0$. Clearly, $\rho$ as given by Eq.~(\ref{rho}) may be expressed in 
this form with $c_1=\tilde m^{-1}-2b$, $c_2=\tilde m^{-1}$, $x_1=\sin(\omega t)$ and 
$x_2=\cos(\omega t)$. 

We conclude this article with the following remarks.
	\begin{itemize}
	\item[1.] Our results on cranked Hamiltonian can be easily generalized to the Hamiltonians of the form 
	\be
	H(t)=h(t)e^{-ig(t)K}H_0 e^{ig(t)K}\;,
	\label{con1}
	\ee
where $g$ and $h$ are real-valued functions and $K$ and $H_0$ are Hermitian operators. 
	\item[2.] The analogy between the generalized harmonic oscillator and the interaction 
of a spinning particle with a changing magnetic field \cite{nova} suggest that we can repeat 
our analysis for the latter system and construct nonstationary cyclic states even for the case 
of constant magnetic fields.
	\end{itemize}

\end{document}